\documentclass[sigconf]{acmart}

\AtBeginDocument{%
  }

\copyrightyear{2026}
\acmYear{2026}
\setcopyright{cc}
\setcctype{by}
\acmConference[WWW '26]{Proceedings of the ACM Web Conference 2026}{April 13--17, 2026}{Dubai, United Arab Emirates}
\acmBooktitle{Proceedings of the ACM Web Conference 2026 (WWW '26), April 13--17, 2026, Dubai, United Arab Emirates}
\acmPrice{}
\acmDOI{10.1145/3774904.3792625}
\acmISBN{979-8-4007-2307-0/2026/04}

\usepackage{subcaption} 
\usepackage{mdframed}
\usepackage{fancyvrb}
\usepackage{url}
\usepackage{xspace}
\usepackage{cleveref}
\usepackage{enumitem}
\usepackage{hyperref}
\newcommand{\robots}{\texttt{robots.txt}\xspace}

\begin{document}

\title[A Study of robots.txt Gatekeeping on the Web]{Is Misinformation More Open? \\A Study of robots.txt Gatekeeping on the Web}

\author{Nicolas	Steinacker-Olsztyn}

\affiliation{%
  \institution{Saarland University}
  \city{Saarbr\"ucken}
  \country{Germany}}
  
\email{nsteinac@mpi-inf.mpg.de	}

\author{Devashish Gosain}
\affiliation{%
  \institution{IIT Bombay}
  \city{Mumbai}
  \country{India}}
\email{dgosain@cse.iitb.ac.in}

\author{Ha Dao}
\authornote{Corresponding author.}
\affiliation{%
  \institution{Max Planck Institute for Informatics}
  \city{Saarbr\"ucken}
  \country{Germany}}
\email{hadao@mpi-inf.mpg.de}

\iftrue
 \newcommand{\reviewerFive}[1]{\textcolor{red}{\textbf{[Comment: }\textit{#1} -- Reviewer 5\textbf{]}}}

  \newcommand{\reviewerOne}[1]{\textcolor{red}{\textbf{[Comment: }\textit{#1} -- Reviewer 1\textbf{]}}}

    \newcommand{\reviewerTwo}[1]{\textcolor{red}{\textbf{[Comment: }\textit{#1} -- Reviewer 2\textbf{]}}}

  \newcommand{\reviewerThree}[1]{\textcolor{red}{\textbf{[Comment: }\textit{#1} -- Reviewer 3\textbf{]}}}

\else
    \newcommand{\reviewerFive}[1]{\ignorespaces}
    \newcommand{\reviewerOne}[1]{\ignorespaces}
    \newcommand{\reviewerTwo}[1]{\ignorespaces}
    \newcommand{\reviewerThree}[1]{\ignorespaces}
   
\fi

\newcommand{\todo}[1]{\textcolor{red}{#1}}

\newcommand{\nico}[1]{\textcolor{teal}{#1}}
\newcommand{\hd}[1]{\textcolor{blue}{#1}}

\newcommand{\dg}[1]{\textcolor{magenta}{#1}}

\begin{abstract}

Large Language Models (LLMs) are increasingly relying on web crawling to stay up to date and accurately answer user queries. These crawlers are expected to honor \robots files, which govern automated access. In this study, for the first time, we investigate whether reputable news websites and misinformation sites differ in how they configure these files, particularly in relation to AI crawlers. Analyzing a curated dataset, we find a stark contrast: 60.0\% of reputable sites disallow at least one AI crawler, compared to just 9.1\% of misinformation sites in their \robots files. Reputable sites forbid an average of 15.5 AI user agents, while misinformation sites prohibit fewer than one. 
We then measure active blocking behavior, where websites refuse to return content when HTTP requests include AI crawler user agents, and disclose that both categories of websites utilize it.
Notably, the behavior of reputable news websites in this regard aligns more closely with their declared \robots directive than with that of misinformation websites.
Finally, our longitudinal analysis reveals that this gap has widened over time, with AI-blocking by reputable websites via \robots increasing from 23\% in September 2023 to nearly 60\% by May 2025. 
Our findings suggest a growing asymmetry in content accessibility, as reflected in \robots directives, that may shape the training data available to LLMs, raising essential questions for web transparency, data ethics, and the future of AI training practices.


\end{abstract}




\authorsaddresses{}

\maketitle

\newcommand\webconfavailabilityurl{https://doi.org/xxxx}
\ifdefempty{\webconfavailabilityurl}{}{
\begingroup\small\noindent\raggedright\textbf{Resource Availability:}
The dataset and analysis scripts of this paper have been made publicly available at \url{https://doi.org/10.17617/3.WW04D8}.
\endgroup
}

\section{Introduction}
Web scraping, the automated process of extracting information from websites, has long played a foundational role in the Internet ecosystem~\cite{gray1995measuring}. It supports services such as search engine indexing, price comparison tools, and competitive intelligence. More recently, it has become a core component in the development of large-scale generative AI models. 
These Large Language Models (LLMs) require enormous volumes of training data, with size being a key variable for model performance ~\cite{kaplan2020scaling, villalobos2024}. To this end, the public web remains a low-cost, attractive source. Major model developers, including those behind OpenAI's Chat-GPT~\cite{GPTdoc}, Google's Bard (now known as Gemini) \& Vertex AI~\cite{googleExtended}, 
and Anthropic's Claude~\cite{anthropicClaude}, openly acknowledge the use of web scraping to construct their training corpora~\cite{abdin2024phi,brown2020language,chowdhery2023palm,grattafiori2024llama,team2024jamba,touvron2023llama}.

This practice has intensified concerns around the quality of scraped content, particularly in the context of misinformation. Misinformation, broadly defined as false or misleading content~\cite{lazer2018science}, has long existed, but the
Internet and social media have amplified their reach and influence.
 It has been linked to real-world harms in public health, elections, and economic stability~\cite{who2021misinformation,bastos2019brexit,roberts2017hoax}. Unlike traditional media, many misinformation websites operate without editorial oversight and are optimized to maximize engagement rather than accuracy~\cite{papadogiannakis2023funds}. 
If LLMs rely more heavily on content from misinformation websites, this could affect the quality and reliability of their outputs. While it is difficult to isolate the influence of individual sources, concerns about training data quality are substantiated by independent audits. A 2025 NewsGuard report~\cite{NewsGuardMayAI} found that 28\% of responses from 11 leading LLMs contained false information when prompted with known misinformation claims. Similarly, a 2023 analysis of Google's C4 dataset, used by models from Google and Meta, identified content tied to white supremacy, propaganda, and conspiracy theories. The authors also noted a lack of transparency about the sources included in many training datasets~\cite{insideWP}. These dynamics underscore the importance of understanding how reputable news and misinformation sources signal their preferences for AI-related access.
Thus, in this paper, we conduct the first longitudinal analysis of how reputable and misinformation websites utilize \texttt{robots.txt} directives, as defined by the Robots Exclusion Protocol (REP) \cite{rfc9309}, a voluntary mechanism for signaling crawler access preferences to regulate interactions with AI user agents. 
This analysis is especially relevant given recent findings that AI crawlers claiming to respect \robots generally comply with these directives~\cite{liu2025somesite}, making exclusion rules a practical tool for managing access to web content.


Our approach involves first curating the list of websites for analysis. For this, we utilize classifications from Media Bias/Fact Check (MBFC) \cite{mbfc}, which categorizes news sources based on their credibility and factual reporting (\autoref{subsec:website_list}). We obtained a total of 4,079 websites, consisting of 3,369 reputable and 710 misinformation websites. We then identified and curated a comprehensive set of 63 AI user agents (e.g., CCBot, ChatGPT-User, Claude-Bot, and Google-Extended). 
Finally, we crawl these websites from seven geographically distributed vantage points, collecting data, such as responses and \robots, to determine behavior regarding AI agents (\autoref{subsec:Robots.txt_crawling}).
We also conducted a longitudinal analysis by accessing historical snapshots from the Internet Archive to observe exclusion trends. We summarize our main findings as follows:

\begin{itemize}[leftmargin=4mm]
    \item Our analysis shows that reputable news websites are substantially more likely to configure \robots directives targeting AI crawlers compared to misinformation websites. Among websites with a \robots file, 60.0\% of reputable sites include a \texttt{DisallowAll} directive for at least one AI agent, compared to just 9.1\% of misinformation sites. On average, reputable news websites reference 15.5 distinct AI agents, while misinformation websites reference only 0.77 (\autoref{subsec:results_prevalence}). 
    
    \item 
    We find that reputable news websites are more likely to include AI-specific directives in their \robots files and tend to specify more extensive and targeted restrictions on automated crawlers.
    Over 50\% disallow GPTBot, and 40–50\% restrict other major AI crawlers such as CCBot, ClaudeBot, ChatGPT-User, and Google-Extended. In contrast, misinformation websites rarely prohibit any specific AI agent, with \texttt{DisallowAll} rates remaining below 5\% across all cases. Reputable sites also maintain broader exclusion lists: 25\% disallow more than 10 AI agents, and the most restrictive restricts up to 54 distinct agents. Meanwhile, over 80\% of misinformation websites do not disallow a single AI agent (\autoref{subsec:results_behavior}).

    \item We measure the adoption of active blocking mechanisms that prevent AI agent crawlers from accessing websites, finding that both misinformation and reputable news websites utilize active blocking to restrict access from AI crawlers. This is found to be performed purely at the level of the declared \texttt{User-agent} header. For instance, 16.9\% of the studied reputable news websites were found to block both ClaudeBot and Anthropic-AI, compared to 9.8\% of misinformation websites. Nearly half of the reputable news websites in this category also had corresponding \robots rules disallowing these crawlers access, compared to less than 2\% for misinformation websites (\autoref{subsec:activeblocking}).
    \item By using six snapshots collected between September 2023 and May 2025 from the Internet Archive, we observe that reputable news websites are rapidly and systematically responding to the rise of AI crawlers by expanding and updating their \robots directives. The percentage of reputable sites disallowing at least one AI agent increased from 23\% to nearly 60\%, with the median number of disallowed agents exceeding 25 by early 2025. In contrast, misinformation websites remain largely passive in updating their \robots directives: their disallow rates stay below 10\%, and their \robots include only a few agents (\autoref{subsec:results_overtime}).
    
\end{itemize}

We discuss the implications of our findings, limitations, and ethical considerations in \autoref{sec:Discussion}.
\section{Background and Related Work}
In this section, we explain the utility and structure of \robots, followed by the related work concerning \robots.

\subsection{Robots.txt}
The Robots Exclusion Protocol (REP) was introduced in 1994 as a practical mechanism to manage automated access to websites~\cite{1994robots}. As web crawlers became increasingly common on the early Web, website owners needed a way to distinguish between cooperative bots and those exhibiting disruptive or abusive behavior. Among the most widely adopted anti-scraping solutions, the REP allows webmasters to publish a \robots file at the root of their domain to define access policies for web crawlers~\cite{rfc9309}. These rules specify which bots are permitted to access the site, which subdomains or paths they may crawl, and how long they must wait between successive requests.
While adherence to these directives is voluntary and relies on crawler compliance, the REP has become a de facto standard for managing bot behavior on the Web.

\autoref{tab:robots-fields} summarizes the most commonly used fields in \robots files. Each set of rules begins with the \texttt{User-agent} directive, which identifies the bot to which the rules apply. Access control is defined by using the \texttt{allow} and \texttt{disallow} directives, specifying URL paths that a bot is explicitly permitted or forbidden to crawl, respectively. To mitigate excessive load on servers, the \texttt{crawl-delay} directive sets a minimum interval between successive requests made by a bot. Finally, the \texttt{sitemap} field provides the location of a sitemap file, which offers a structured overview of the URLs intended for crawling.

\begin{table*}[t!]
\centering
\caption{Common fields in \robots files. The \robots directives define how bots are allowed to interact with a website.}
\begin{tabular}{lr}
\toprule
\textbf{robots.txt field} & \textbf{Description} \\
\midrule
\texttt{User-agent}  & Identifies the bot using its declared \texttt{User-agent} string (e.g., \texttt{Googlebot}). \\
\texttt{allow}       & Specifies URLs that the identified bot is permitted to access. \\
\texttt{disallow}    & Specifies URLs that the identified bot is not permitted to access. \\
\texttt{crawl-delay} & Defines the minimum delay (in seconds) between successive requests by the bot. \\
\texttt{sitemap}     & Indicates the location of the site's sitemap containing URLs for crawling. \\
\bottomrule
\end{tabular}
\label{tab:robots-fields}
\end{table*}




\subsection{Related Work}
The REP has been studied for decades as a voluntary mechanism for websites to control automated access. 
Sun et al.~\cite{sun2007large} conducted one of the first large-scale studies of \texttt{robots.txt} adoption, analyzing 7,593 websites across education, government, news, and business sectors. Their longitudinal analysis showed that the use of \texttt{robots.txt} increased over time.
Sun et al.~\cite{sun2007determining} analyzed \texttt{robots.txt} files from over 7,500 websites and found that popular search engine crawlers like Google and Yahoo were more often allowed access. They showed that this bias correlates with market share, suggesting that well-known crawlers get preferential treatment on the Web.
Kolay et al.~\cite{kolay2008larger} conducted a large-scale analysis of over 2.2 million \texttt{robots.txt} files to investigate potential bias in crawler access permissions. Building on prior findings that suggested crawler-specific bias correlating with search engine market share, they replicated and extended this analysis using data from high-ranking and randomly selected sites. While they confirmed that some sites exhibited crawler-specific restrictions, their main finding was that the top two search engine crawlers generally had equal access to content.
 These foundational studies established that REP was widely adopted and revealed patterns of preferential access for dominant search engines. However, they predated the emergence of AI-specific crawlers and the rise of web scraping for large-scale language model training.

Recent work has examined the use of \robots in the context of AI, focusing on their role in signaling consent and restricting data access. Kim et al.~\cite{kim2025scrapers} found that AI bots show moderate compliance with \robots, with some non-compliance due to spoofing. Longpre et al.~\cite{longpre2024consent} audited 14,000 websites and reported a rise in AI-specific restrictions, along with inconsistencies between stated policies and technical enforcement. Chang et al.~\cite{chang2025liabilitiesrobotstxt} explored the legal status of \robots, arguing that their violations may carry legal, not just ethical, consequences. Jiménez et al.~\cite{jimenez2024ai} emphasized the protocol's limitations for expressing data preferences in modern AI contexts. Liu et al.~\cite{liu2025somesite} found that while adoption of newer signals like \emph{NoAI} remains low, major AI crawlers generally respect \robots. They also studied network level crawling blocks provided by Cloudflare, and the prevalance of active blocking based off of the declared \texttt{User-agent} header. Dinzinger et al.~\cite{dinzinger2024} tracked longitudinal changes in content control, noting growing use of exclusion mechanisms in response to AI scraping. 
These studies highlight both the growing use of AI-specific exclusions and the limitations of existing protocols in the face of evolving data practices. 
Our work extends this line of research by examining how websites with differing credibility, reputable and misinformation, use the \robots to control access by AI-related user agents.



\section{Methodology}
\begin{figure*}
    \centering
    \includegraphics[width=0.73\textwidth]{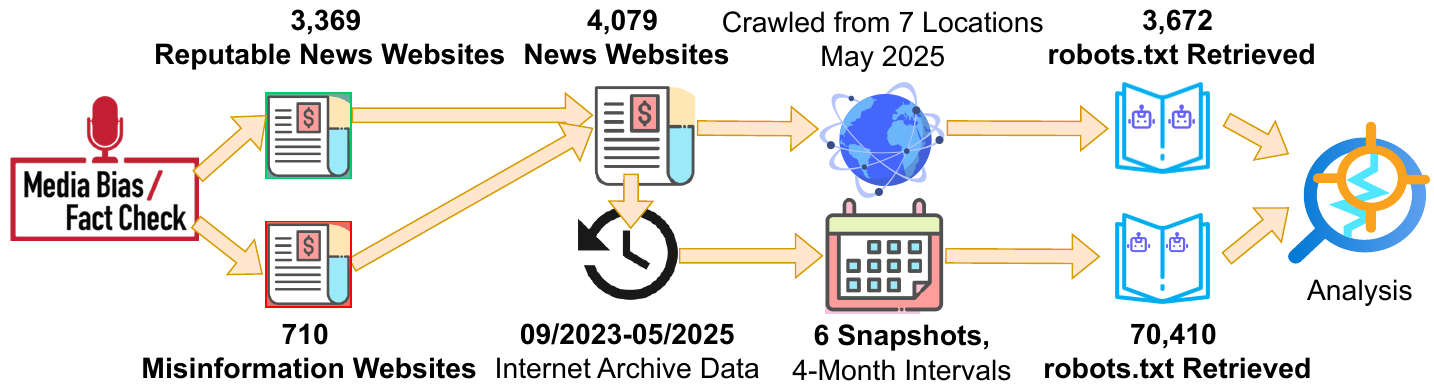}
    \caption{Methodology overview.}
    \label{fig:overview}
\end{figure*}

In this section, we describe the methodology used to collect, process, and analyze \robots files from reputable news and misinformation websites. Our approach consists of three components: constructing website lists based on credibility, retrieving and validating \robots files across multiple geographic vantage points and historical snapshots from the Internet Archive, and compiling a curated list of AI-related user agents (see~\autoref{fig:overview}).

\subsection{Reputable News and Misinformation Website Lists}
\label{subsec:website_list}


Similar to prior research that uses Media Bias/Fact Check (MBFC)~\cite{mbfc} to classify the credibility and bias of news domains~\cite{chen2020proactive,chowdhury2020joint,ghanem2021fakeflow,gruppi2021nela,zhou2020recovery,papadogiannakis2023funds}, we rely on MBFC to construct two domain sets: misinformation websites and reputable news websites. MBFC is an independent organization that evaluates the political bias and factual accuracy of online media sources through a transparent, standardized manual methodology~\cite{mbfc-methodology}.

\noindent \textbf{Misinformation websites:}
We select websites labeled by MBFC as either \emph{Questionable Source} or \emph{Conspiracy-Pseudoscience}, rated as having \emph{Low} or \emph{Very Low} factual reporting and described as having \emph{Low} overall credibility. According to MBFC, such websites frequently promote false or misleading narratives, lack credible sourcing, and exhibit minimal editorial transparency. This selection resulted in a list of \textbf{710} unique misinformation websites.

\noindent \textbf{Reputable news websites:}
As a comparison group, we include websites labeled by MBFC as \emph{Pro-Science}, \emph{Least Biased}, \emph{Left-Center}, or \emph{Right-Center}, all rated as having \emph{High} or \emph{Very High} factual reporting and \emph{High} overall credibility. 
These labels were selected as they return websites that typically adhere to journalistic standards and provide evidence-based reporting \cite{papadogiannakis2023funds}. This yielded \textbf{3,369} unique reputable news websites. 


To understand the popularity of the sites in our dataset, 
we show their ranking based on the Tranco list~\cite{tranco} in~\autoref{fig:rank_1} in the Appendix.
We find that both reputable and misinformation websites can have very high popularity, including 15 misinformation websites among the Top 10K most visited domains.





\subsection{Robots.txt Collection}
\label{subsec:Robots.txt_crawling}


Building on the domain lists described in~\autoref{subsec:website_list}, we collected \robots files for both reputable news and misinformation websites. 
To conduct our study responsibly, we first defined a custom \texttt{User-agent} string that included a contact email address. 
For each domain, we issued an HTTP request to the standard root location of the \robots file (e.g., \url{https://example.com/robots.txt}), following the REP convention. 
In parallel, we also fetched the homepage of each domain. This dual-request setup allowed us to determine whether the \robots file was genuinely unavailable or if the site itself was non-functional. If the \robots file was inaccessible, we analyzed the status codes and response behavior of both requests to distinguish between these cases. We recorded the HTTP status codes of both the homepage and \robots responses, along with any URL redirections observed during the retrieval process. In addition, 
to improve measurement robustness and account for potential geo-blocking or region-specific bot responses, we deployed AWS cloud instances across seven geographic regions: one vantage point in Germany, one in Sweden, one in the United States, one in Brazil, and one each in Africa, Asia, and Australia. Each site was crawled three times, and we retained the most permissive valid result.
We consider a \robots file valid if it returns an HTTP 2XX response and contains parseable plain text content.
Responses served as HTML are automatically excluded.
The crawler would also allow for up to 3 redirections if they were within the same registered domain and resulted in a \robots file. Other responses and status codes, such as in the 4XX or 5XX range, were marked as unsuccessful. It should be noted that status codes in the 4XX range are defined in the REP standard~\cite{rfc9309} to permit crawlers to assume no access restrictions. Using this methodology, we determined 3,299 reputable news websites whose homepages were still operational. Among these, 3,179 served a valid \robots file, resulting in an adoption rate of 96.4\%. For misinformation websites, 672 were operational, of which 493 (73.8\%) provided a valid \robots file.

In addition, to examine how the use of AI agent-specific directives in \robots has evolved over time, we conduct a \textbf{longitudinal analysis} using data from the Internet Archive---an online digital library that preserves and provides access to historical snapshots of websites~\cite{internetarchive2021about}.
Our dataset comprises six snapshots of \robots files and associated web pages, collected at 4-month intervals between 1 September, 2023 and 1 May, 2025.\footnote{We excluded our own crawl data from this analysis, as our crawlers use different \texttt{User-agent} headers than those used by the Internet Archive. To ensure a fair and consistent comparison across snapshots, we relied solely on data from the Internet Archive for the longitudinal analysis.}
The initial date was chosen to assess changes to how reputable news websites and misinformation websites responded one month after the announcement of OpenAI's GPTBot crawler, which received notable backlash upon its release \cite{guardianGPT}. For each snapshot, the most recently available \robots for a website that occurred before the date in question was selected.

\subsection{AI Agents Collection}
\label{subsec:AI-agents_collection}

To assess the prevalence of AI-related crawlers in our dataset, we first identified which \texttt{User-agent} strings were associated with AI services. 
To this end, we relied on four distinct sources:
(1) the website Dark Visitors, which provides a comprehensive catalog of known web crawlers, including those linked to AI tools~\cite{darkvisitors};
(2) Cloudflare’s Radar platform, which offers real-time insights into web agent activity and explicitly labels several AI-associated crawlers~\cite{cloudflareRadar};
(3) a prior study on AI crawler behavior by Liu et al. \cite{liu2025somesite}, which supplemented the above sources with several AI \texttt{User-agent} identifiers; and
(4) \texttt{ai.robots.txt}, a public repository that maintains a curated list of AI-focused user agents \cite{airobotstxt}.

Combining these sources, we manually verified and compiled a final list of 63 unique AI user agents for our analysis. Notably, this list includes crawlers like CCBot, which is not operated by an AI company, but is frequently used by them due to the public availability of its crawled data~\cite{commonCrawl}.

\section{Results}
In this section, we analyze \robots directives targeting AI crawlers across reputable and misinformation websites.
We assess the prevalence and configuration of AI-specific exclusions, identify the most frequently blocked agents, measure active blocking of AI crawlers, and examine how these practices have changed over time using six Internet Archive snapshots.

\subsection{Prevalence of AI Agents-specific Directives}
\label{subsec:results_prevalence}
To understand how websites handle access from AI-specific user agents, we first examine the prevalence and configuration of \robots directives across reputable news websites and misinformation websites in~\autoref{tab:robotsstats}.
Overall, we find that directives targeting AI agent-specific user agents are widely adopted among reputable news websites but remain uncommon among misinformation sites. 
In our dataset, 96.4\% of reputable news websites and 73.8\% of misinformation websites include a \robots file. 
Among these, 61.3\% of reputable news websites explicitly mention at least one AI agent, while only 12.3\% of misinformation sites do so. Reputable news sites also frequently implement restrictive policies: 60.0\% employ a \texttt{DisallowAll} directive for at least one AI agent, compared to just 9.1\% among misinformation sites.

Additionally, reputable news websites mention a much larger number of distinct AI agents on average $15.5 (\pm 12.6)$, compared to misinformation sites' \(0.77(\pm 3.2\)). 
A partially allowed agent is both defined and given at least one path that it cannot access.
Interestingly, misinformation websites have double the percentage of partially allowed AI agents, with five different misinformation websites in our dataset explicitly allowing partial access to 16 different AI agents. Through our analysis of the retrieved files, we note that despite \robots offering flexibility for what directories and paths a crawler may access, when an AI agent is defined, nearly all websites opt to fully restrict them. 98.4\% of AI agents once declared on reputable news websites are subsequently entirely restricted via \texttt{DisallowAll}, with 72.7\% of misinformation websites also doing so. For the remaining AI agents that explicitly have access, both categories of websites share a nearly identical ratio of 83\% and 17\% for partial allowance and full allowance, respectively.

Given that reputable news information websites are more prevalent in the Tranco Top 1M, it is reasonable to infer that the observed differences may not stem from the nature of their content itself (i.e., whether it is reputable or misinformation) but from their popularity. More popular websites are likely to possess greater resources and a more substantial commitment to employing various tools for access management.

\begin{figure}
    \centering
    \includegraphics[width=0.8\linewidth]{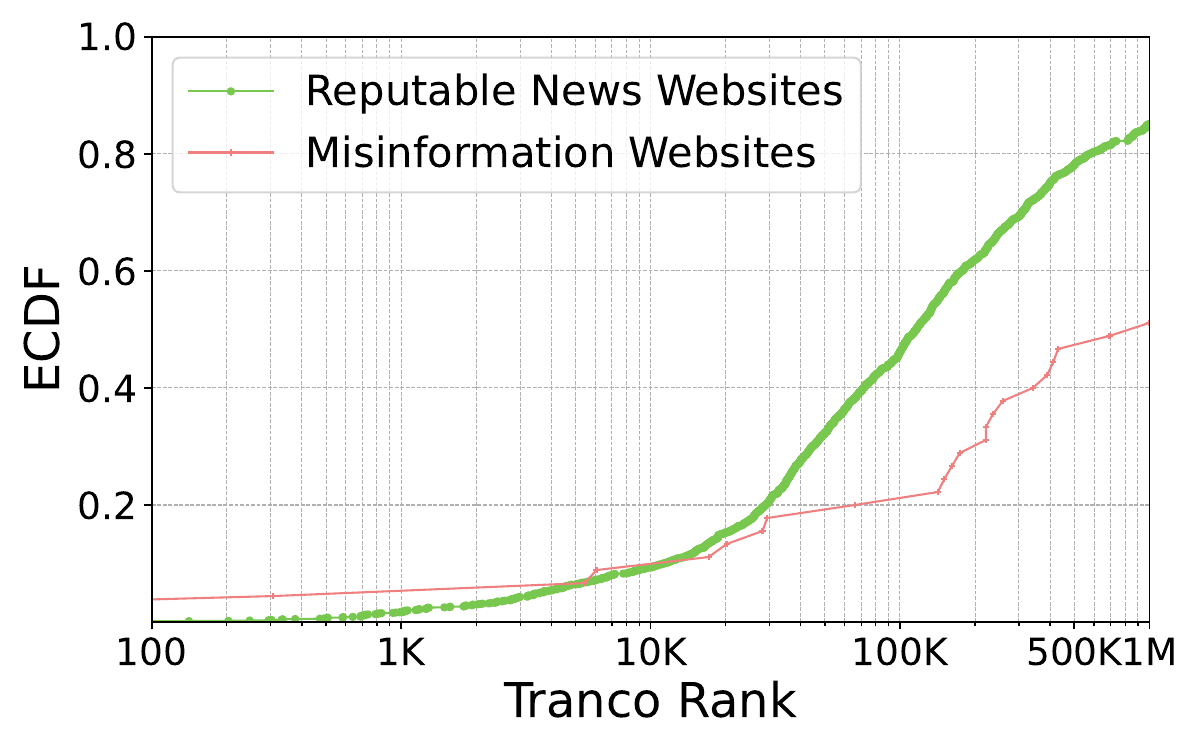}
    \caption{Tranco popularity for websites that have a \texttt{DisallowAll} rule for at least one AI agent crawler. }
    \label{fig:rankatleast}
\end{figure}

To measure this, we examined the popularity of reputable news websites and misinformation websites that included a \texttt{DisallowAll} rule for at least one AI agent (45 misinformation websites and 1,908 reputable news websites, see~\autoref{fig:rankatleast}). 
While a similar percentage of both categories align for roughly the first 10,000 websites, a clear separation of practices occurs afterward. This indicates that popularity alone does not determine the \robots practices of news websites.

Furthermore, we manually inspected the retrieved \robots files of all misinformation websites ranked within the Tranco Top 10k. We identified 15 misinformation websites in the Tranco Top 10k, but one was inaccessible and therefore excluded from further evaluation.
For the remainder, 13 out of 14 websites, served a \robots file, while one returned a 404 response. 
Of the 13 websites that provided a \robots file, 9 did not include a single \texttt{DisallowAll} directive targeting any AI agent. Since a 404 response indicates that a crawler may assume no explicit restrictions, this means that 10 out of 14 (71.4\%) of the most popular misinformation websites' \robots do not impose any limitations on AI crawling access.


\vspace{2mm}
\begin{mdframed}[linecolor=cyan!60!black, backgroundcolor=cyan!5]
\textbf{Key takeaway:} 
Reputable news websites are substantially more likely to restrict access to AI crawlers via \robots, whereas misinformation sites rarely take similar measures. 
\end{mdframed}

\begin{table}[!t]
    \centering
    \caption{Basic statistics on \robots files.}
    \resizebox{\linewidth}{!}{%
    \begin{tabular}{l|rr}
    \toprule
        \textbf{Metric}  & \textbf{Reputable News} & \textbf{Misinformation} \\
    \midrule
    \textbf{\# Success websites} & 3,299 & 672\\
         \textbf{Adoption among websites} & 3,179 (96.4\%) & 493 (73.8\%) \\

          \textbf{DisallowAll for $\geq$1 AI agent} & 60.0\% & 9.1\% \\
          \textbf{Partial Disallow for $\geq$1 AI agent} & 0.7\% & 1.4\% \\
          \textbf{Avg. \# of AI agents} & \(15.5 (\pm 12.6\)) & \(0.77(\pm 3.2\)) \\
          



    \bottomrule
    \end{tabular}
    }
    \label{tab:robotsstats}
\end{table}


\subsection{Restricted AI Agents}

\label{subsec:results_behavior}
\begin{figure*}[htbp]
    \centering
    \begin{minipage}[t]{0.45\linewidth}
        \centering
        \includegraphics[width=\linewidth]{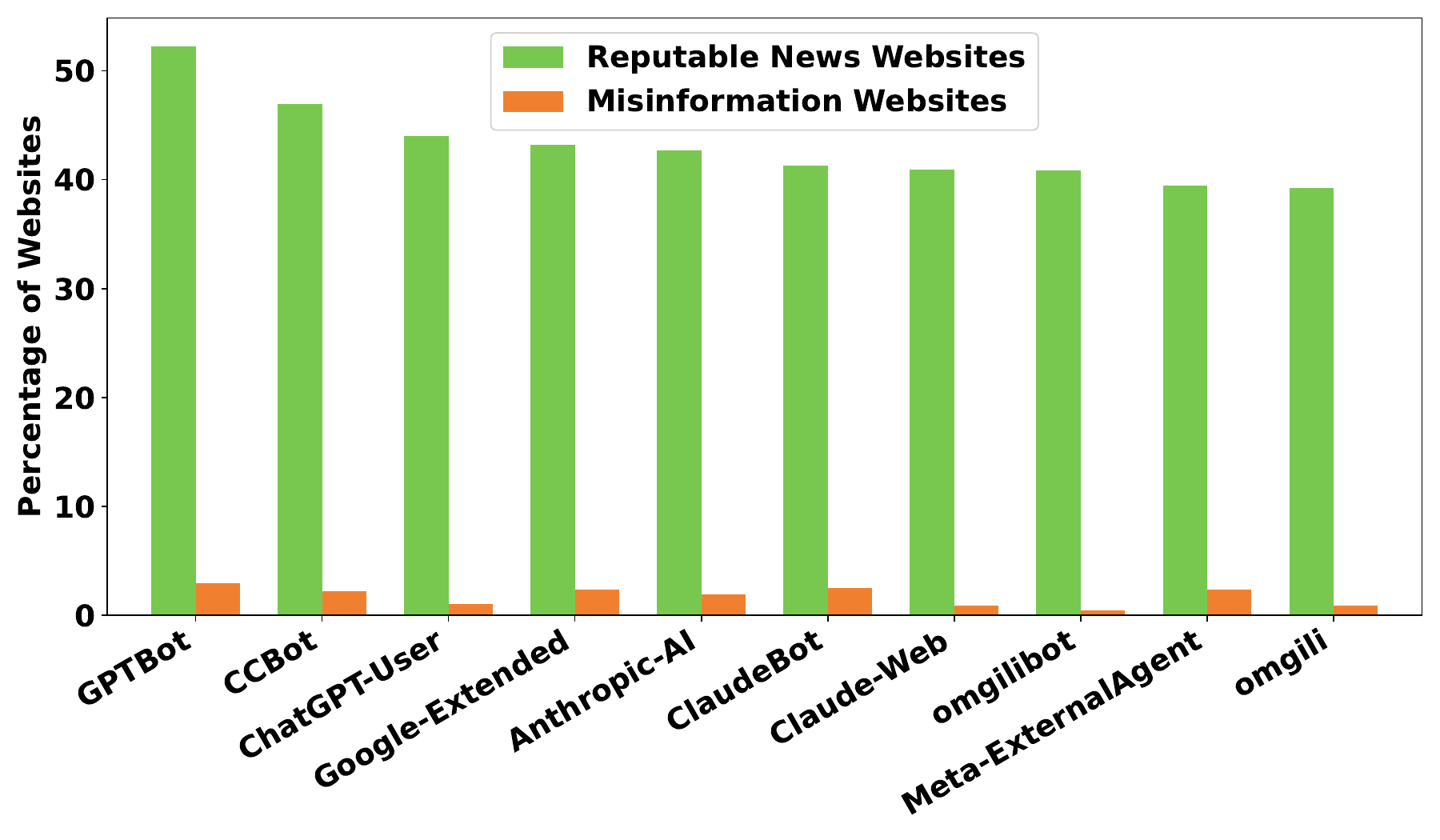}
        \caption{\texttt{DisallowAll} usage in \robots for the Top 10 AI agents in reputable news and misinformation sites.}
    
        \label{fig:top20AIAgents_v2}
    \end{minipage}%
    \hfill
    \begin{minipage}[t]{0.45\linewidth}
        \centering
        \includegraphics[width=\linewidth]{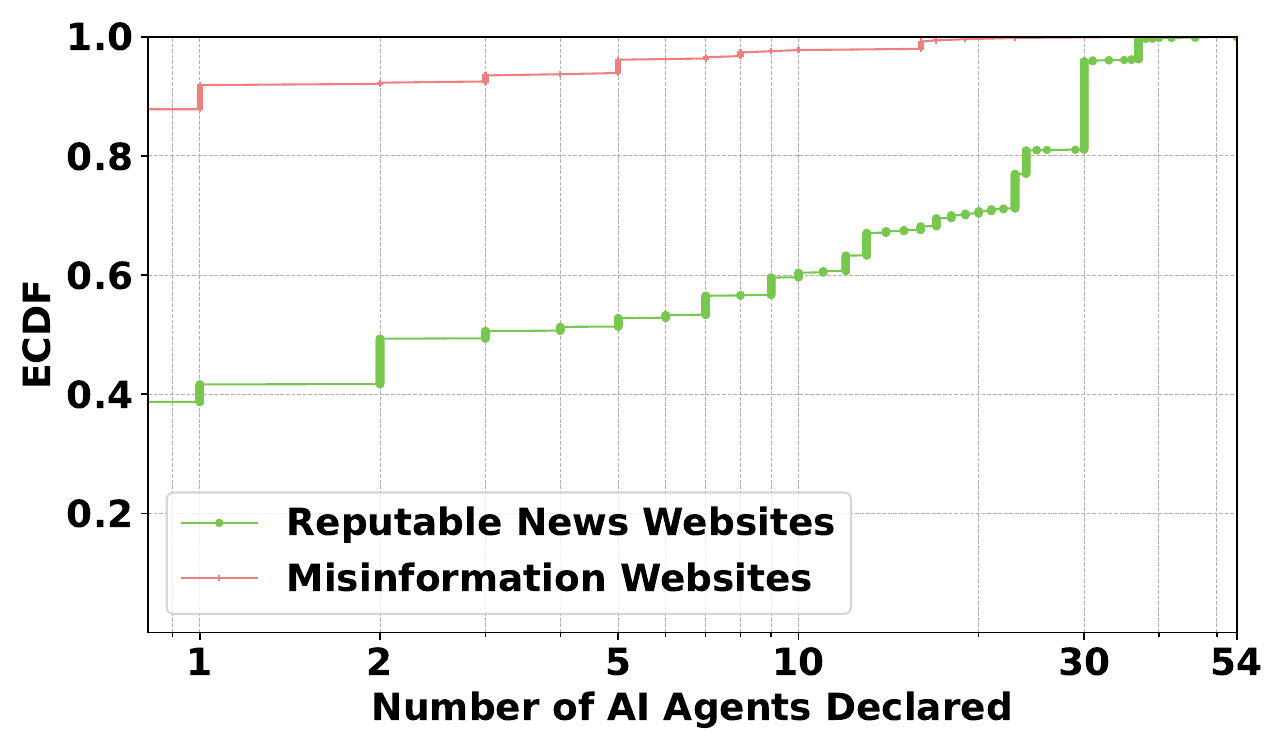}
        \caption{Number of AI agents disallowed per website in \robots on reputable news and misinformation sites. 
        }
        \label{fig:ecdf_AI_agent_stats}
    \end{minipage}
\end{figure*}

 We now present the percentage of reputable news and misinformation websites that apply \texttt{DisallowAll} directives to the top 20 AI user agents in their \robots files in~\autoref{fig:top20AIAgents_v2}. Reputable news websites exhibit consistent and selective restriction behavior, with over 50\% disallowing GPTBot, and around 40–-50\% discouraging other prominent AI crawlers such as CCBot, ChatGPT-User, ClaudeBot, and Google-Extended. In contrast, misinformation websites rarely disallow any individual AI agents, with \texttt{DisallowAll} rates remaining below 5\% across all cases.

We further report the distribution of the number of AI agents disallowed per website in \robots across reputable news and misinformation websites in~\Cref{fig:ecdf_AI_agent_stats}. 
Among reputable news websites, approximately 40\% do not disallow any AI agents, while the remaining apply increasingly extensive restrictions. Around 25\% of reputable sites disallow more than 10 agents, and the three most restrictive sites\footnote{\texttt{arkansasonline.com}, \texttt{timesfreepress.com}, and \texttt{nwaonline.com}} each disallow access to 54 distinct AI user agents. In contrast, misinformation websites display a much narrower range: over 90\% disallow two or fewer agents, with more than 80\% disallowing none.

\vspace{2mm}
\begin{mdframed}[linecolor=cyan!60!black, backgroundcolor=cyan!5]
\textbf{Key takeaway:} 
 Reputable news websites not only adopt AI-related directives more frequently, but also tailor their exclusions to specific user agents.
  In addition, there is a clear disparity in exclusion practices, with reputable news websites more actively and comprehensively restricting access from AI crawlers via \robots.
\end{mdframed}

\subsection{Active Blocking of AI Crawlers}
\label{subsec:activeblocking}
Active blocking refers to a content access control technique in which access to a website's content can be denied based on specific parameters. While \robots are the de facto standard for content control, there have been reported incidents of AI crawlers not respecting \robots directives in the past \cite{perplexityNYT, perplexityPCMag}. As a result, some websites may likely take this more active approach towards dealing with AI agents, which may not be captured by assessing \robots directives alone. Therefore, to measure the prevalence of active blocking, we adopted a similar methodology as \cite{liu2025somesite} to determine the degree to which websites were taking additional measures against AI crawlers. For this measurement, we retrieved the status of every homepage from our list of reputable and misinformation news websites, recording values such as the returned status code and response size in bytes. There already exists literature to suggest that websites alter both their content and returned status simply to changes in the value of the \texttt{User-agent} string \cite{Santos2016, liu2025somesite}. However, to our knowledge, the active blocking behavior of news websites has never been studied regarding AI agents, offering potential novel insights.

These crawls were conducted sequentially from a single location to remove unintended variance, as prior work has shown that the IP’s geographic location can also affect the response type received \cite{Santos2016}. We had also determined that there was no discernible difference in \robots between regions from our most recent crawl. Furthermore, we chose to utilize simple HTTP requests and a limited set of headers. Since this pattern of requests does not resemble typical user behavior, it allows us to filter out potential non-200 responses that would arise purely from automated detection. To further identify cases where a website’s behavior may have changed for reasons unrelated to the user agent, we performed an additional, final control crawl with the same parameters as the initial one after finishing all other crawls. Websites in the final control crawl that received a different response than the initial control crawl were filtered out from analysis.

With this setup and the curated list of news websites, our crawler initiated requests beginning with the control crawl, which used a custom-defined \texttt{User-agent} header. For the AI agent crawls, the only parameter varied was the \texttt{User-agent} value. Following the approach in \cite{liu2025somesite}, this value was modified to emulate the identifiers of ClaudeBot and Anthropic-AI, two of the most commonly blocked AI crawlers across both misinformation and reputable news websites (see \autoref{fig:top20AIAgents_v2}).
In addition, neither of these agents has publicly available IP address ranges~\cite{darkvisitors}; therefore, site operators are more likely to block them based on their \texttt{User-agent} identifiers.


Our initial control-agent crawl successfully retrieved responses from 2,877 reputable news websites and 576 misinformation websites, as determined by receiving a 200-level status code after up to three potential redirections. A total of 18 websites were excluded at this stage, as they returned 200 responses but were no longer operational (e.g., domains listed for sale). During the final control run, an additional 14 reputable and 11 misinformation domains failed to return a valid 200-level status code and were consequently excluded from further analysis.

When examining the responses towards Anthropic-AI, 516 reputable news websites (18.0\%) appear to perform active blocking against this crawler, compared to 58 for misinformation websites (10.2\% of all misinformation websites in the control group). For ClaudeBot, 711 reputable news websites appear to perform active blocking (24.8\%), compared to 147 misinformation websites (25.6\%). Of these websites, 486 (16.9\%) reputable news websites actively block both Anthropic-AI and ClaudeBot, compared to 56 (9.8\%) misinformation websites that block both agents.

When inspecting the retrieved \robots for these two categories of websites, 230 of the reputable news websites have \texttt{DisallowAll} for both of these agents (47.3\% of websites that actively block both, and 8\% of all reputable news websites), with 4 only having a \texttt{DisallowAll} rule for Anthropic-AI and 2 having a \texttt{DisallowAll} rule only for ClaudeBot. For misinformation websites, only 1 website that actively blocked both had a \robots rule also disallowing both of these agents access. There were no instances of both agents being actively blocked, with only one being specified in the \robots. 


Although less adopted than \robots, our findings indicate that among reputable news websites, \robots rules correlate with the observed behavior toward AI crawlers, as determined by their declared \texttt{User-agent} strings. In contrast, while the adoption of \robots and the rules for AI agents are less prevalent among misinformation websites, these websites appear to rely relatively more frequently on active blocking mechanisms. 
As noted in~\cite{liu2025somesite}, although active blocking may appear to be a strictly superior approach, it cannot fully replace the use of directives in \texttt{robots.txt}. Active blocking is inherently an all-or-nothing mechanism that can lead to unintended consequences, such as preventing other legitimate activities (e.g., web search indexing) conducted by the same company.
Therefore, reputable news sites benefit from maintaining a combination of directive-based and active blocking strategies, ensuring both fine-grained control over crawler behavior and alignment with evolving AI access policies.

\vspace{2mm}
\begin{mdframed}[linecolor=cyan!60!black, backgroundcolor=cyan!5]
\textbf{Key takeaway: } 
Both misinformation and reputable news websites utilize active blocking as a mechanism for restricting access from AI crawlers, as determined by their declared \texttt{User-agent} string. The behavior of reputable news websites in this regard aligns more closely with their declared \robots directive than misinformation websites. 

\end{mdframed}













\subsection{Longitudinal Analysis of AI-Specific \robots Directives}
\label{subsec:results_overtime}


\begin{figure*}[t!]
    \centering
    \begin{minipage}[t]{0.46\linewidth}
        \centering
        \includegraphics[width=\linewidth]{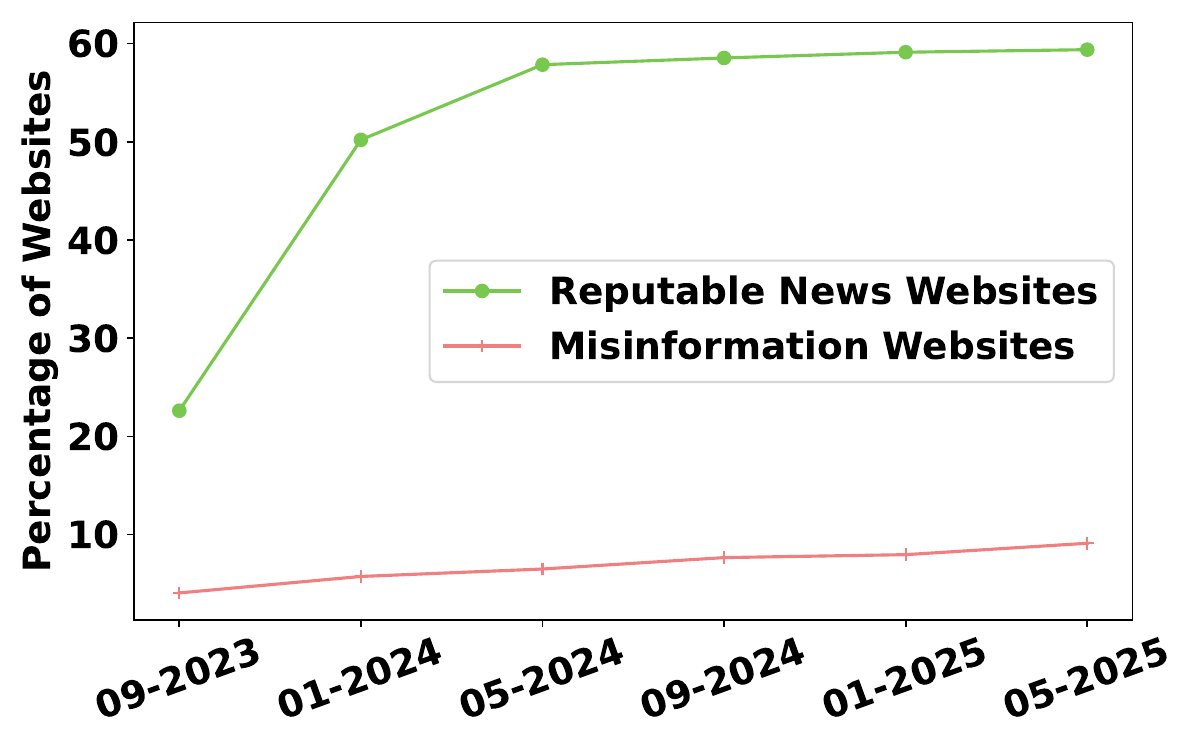}
        \caption{Percentage of sites that \texttt{DisallowAll} at least one AI agent in their \robots over time.
        }
        \label{fig:atleast-one-agent}
    \end{minipage}%
    \hfill
    \begin{minipage}[t]{0.46\linewidth}
        \centering
        \includegraphics[width=\linewidth]{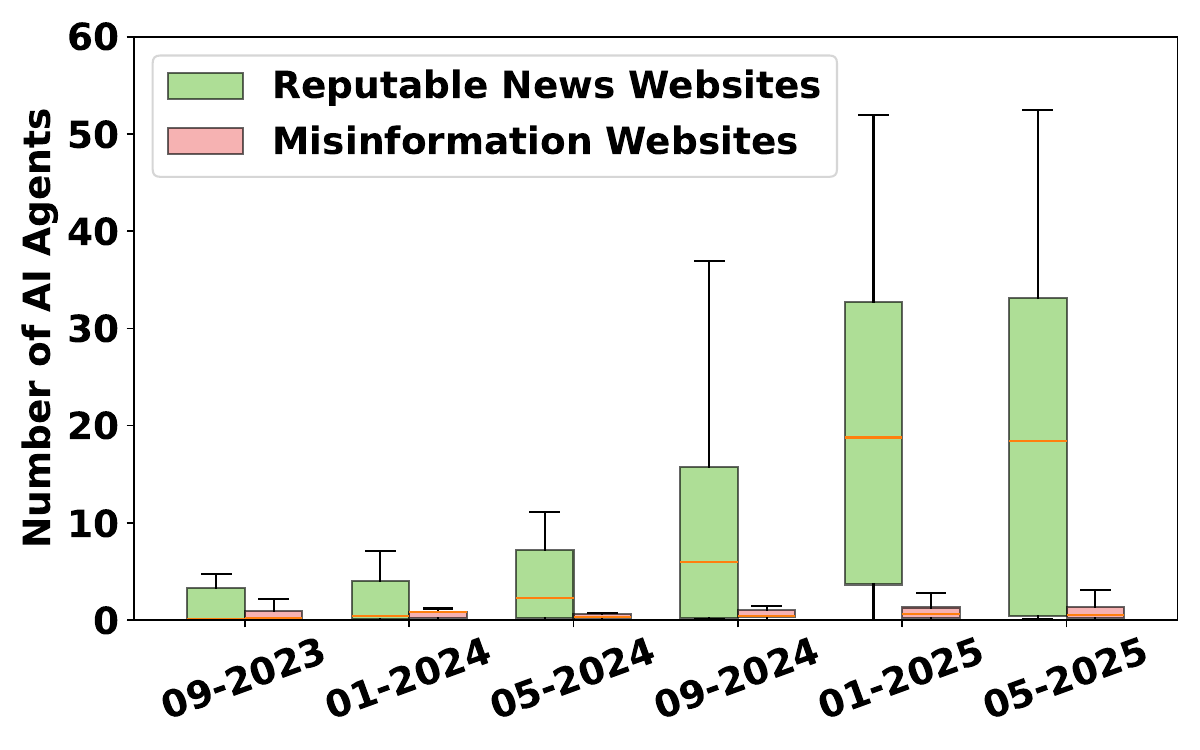}
        \caption{Number of AI agents disallowed per \robots file on reputable news and misinformation sites over time.}
        \label{fig:avg-disallow-ai-agents}
    \end{minipage}
\end{figure*}

We show the evolution of sites disallowing at least one AI agent using a \texttt{DisallowAll} directive, from September 2023 to May 2025 in~\autoref{fig:atleast-one-agent}. We observe a rapid uptake among reputable news websites: the proportion rose from 23\% in September 2023 to 50\% by January 2024 and continued to increase gradually, reaching nearly 60\% by May 2025. This surge aligns temporally with public concerns over the use of copyrighted content in training large language models, as well as with announcements from major AI companies regarding their data-crawling policies~\cite{oecd2025ipai,europarl2025genai}. In contrast, the adoption among misinformation websites remains minimal and increases only modestly, from 4.6\% to 9.2\% over the same period.
Note that, during this period, we observed 27 different reputable news sites and one misinformation site that updated their \robots files to change from previously explicitly allowing an AI agent to explicitly disallowing all access via a \texttt{DisallowAll} directive in their most recent retrieved version. Notably, 26 of these websites moved from granting partial access to applying \texttt{DisallowAll} to PerplexityBot. These rule changes were all made in 07/2024, a month after the crawler made headlines for disobeying \robots directives \cite{perplexityNYT, perplexityPCMag}.
This indicates that reputable news site publishers are informed on recent AI developments and regularly maintain their \robots. Thus, it raises a broader concern: following the existing trend, if reputable websites in the future \textit{deliberately} prohibit AI-related crawlers and misinformation websites do not, it opens the possibility for LLMs and AI-related tools to come into contact with higher ratios of incorrect and biased information.


We then show the distribution of the number of AI agents disallowed per \robots file for reputable news and misinformation websites from September 2023 to May 2025 in~\autoref{fig:avg-disallow-ai-agents}. Among reputable news websites, the median number of disallowed agents steadily increases over time, with a sharp rise beginning in mid-2024. By early 2025, the median exceeds 25 disallowed agents, and the upper quartile reaches nearly 50, indicating that some sites are applying highly comprehensive exclusion lists. The widening interquartile range also suggests growing variation in blocking behavior across reputable domains, likely due to diverging organizational policies or differing adoption timelines.
In contrast, misinformation websites show minimal change throughout the period. The median remains at or near zero, and the entire distribution is tightly concentrated below five disallowed agents. 

Lastly, we track the top 10 AI user agents most frequently disallowed via \robots on reputable and misinformation websites across six time points from September 2023 to May 2025 in~\Cref{tab:top10overtime}.
In the first snapshot, we observe that AI agents are generally represented at relatively low levels in both categories. GPTBot's announcement a month prior, however, had already led to it being by far the most disallowed AI agent by reputable news websites. Interestingly, for almost every snapshot, the most restricted AI agent for misinformation websites was Huawei's PetalBot. This agent only appears in the top 10 of reputable news websites in its first two snapshots. We can also observe that, starting from the 09/2024 snapshot, PerplexityBot appears in the top 10 for the first time and then has its block percentage more than double. This can likely be attributed to the news unfolding of the Amazon-backed service ignoring \robots, and the New York Times subsequently sending Perplexity a cease-and-desist \cite{perplexityNYT, perplexityPCMag}. The increasing pace of reputable news websites disallowance of AI agents in the second half of the snapshots can likely be attributed to major AI organizations making the names and policies of their web crawlers publicly known, and growing focus and scrutiny of AI development as reflected in legislation, such as the EU AI Act which went into effect on 08/2024 \cite{euAIAct}.
Overall, as it pertains to maintaining website access via \robots directives, we observe that reputable sites closely follow developments regarding AI and adapt accordingly, while misinformation websites remain largely passive in this regard.

\definecolor{tab20_0}{RGB}{31,119,180}
\definecolor{tab20_1}{RGB}{174,199,232}
\definecolor{tab20_2}{RGB}{255,127,14}
\definecolor{tab20_3}{RGB}{255,187,120}
\definecolor{tab20_4}{RGB}{44,160,44}
\definecolor{tab20_5}{RGB}{152,223,138}
\definecolor{tab20_6}{RGB}{214,39,40}
\definecolor{tab20_7}{RGB}{255,152,150}
\definecolor{tab20_8}{RGB}{148,103,189}
\definecolor{tab20_9}{RGB}{197,176,213}
\definecolor{tab20_10}{RGB}{140,86,75}
\definecolor{tab20_11}{RGB}{196,156,148}
\definecolor{tab20_12}{RGB}{227,119,194}
\definecolor{tab20_13}{RGB}{247,182,210}
\definecolor{tab20_14}{RGB}{127,127,127}
\definecolor{tab20_15}{RGB}{199,199,199}
\definecolor{tab20_16}{RGB}{188,189,34}
\definecolor{tab20_17}{RGB}{219,219,141}
\definecolor{tab20_18}{RGB}{23,190,207}
\definecolor{tab20_19}{RGB}{158,218,229}

\begin{table*}[t!]
	\centering
	\caption{\texttt{DisallowAll} usage in \robots for the Top 10
		AI agents observed in Reputable News and Misinformation
		sites over time. Each AI agent is shown in a consistent color.
    }
    \label{tab:top10overtime}
	\resizebox{\textwidth}{!}{%
		\begin{tabular}{l|
			rr| rr| rr}
			\toprule
			& \multicolumn{2}{c|}{\textbf{09/2023}}
			& \multicolumn{2}{c|}{\textbf{01/2024}} 
			& \multicolumn{2}{c}{\textbf{05/2024}} \\
			
			\cmidrule(lr){2-3} \cmidrule(lr){4-5} \cmidrule(lr){6-7} 
			\# & \textbf{Reputable News}          & \textbf{Misinformation}          & \textbf{\textbf{Reputable News}}            & \textbf{Misinformation}           & \textbf{Reputable News}            & \textbf{Misinformation}           \\
			\midrule
			1  & \textcolor{tab20_0}{GPTBot} (16.28)\%        & \textcolor{tab20_0}{GPTBot} (2.19\%)         & \textcolor{tab20_0}{GPTBot} (45.10\%)          & \textcolor{tab20_2}{PetalBot} (2.99\%)        & \textcolor{tab20_0}{GPTBot} (52.76\%)          & \textcolor{tab20_2}{PetalBot} (3.87\%)       \\
			2  & \textcolor{tab20_1}{CCBot }(4.79\%)          & \textcolor{tab20_2}{PetalBot} (1.46\%)       & \textcolor{tab20_1}{CCBot }(32.79\%)           & \textcolor{tab20_0}{GPTBot} (2.20\%)          & \textcolor{tab20_1}{CCBot }(40.63\%)           & \textcolor{tab20_0}{GPTBot} (1.93\%)         \\
			3  & \textcolor{tab20_13}{omgilibot}(3.89\%)       & \textcolor{tab20_1}{CCBot }(0.97\%)          & \textcolor{tab20_9}{Google-Extended} (31.10\%) & \textcolor{tab20_1}{CCBot }(1.20\%)           & \textcolor{tab20_9}{Google-Extended} (39.32\%) & \textcolor{tab20_17}{SemrushBot-SWA} (1.55\%)  \\
			4  & \textcolor{tab20_2}{PetalBot} (3.44\%)       & \textcolor{tab20_4}{magpie-crawler} (0.24\%) & \textcolor{tab20_8}{ChatGPT-User} (25.26\%)    & \textcolor{tab20_17}{SemrushBot-SWA} (1.00\%)  & \textcolor{tab20_8}{ChatGPT-User} (34.91\%)    & \textcolor{tab20_1}{CCBot }(0.77\%)          \\
			5  & \textcolor{tab20_3}{omgili} (3.30\%)         & \textcolor{tab20_6}{aiHitBot} (0.24\%)       & \textcolor{tab20_13}{omgilibot} (7.14\%)         & \textcolor{tab20_9}{Google-Extended} (0.80\%) & \textcolor{tab20_10}{Anthropic-AI} (11.11\%)    & \textcolor{tab20_9}{Google-Extended} (0.77\%) \\
			6  & \textcolor{tab20_8}{ChatGPT-User} (1.35\%) & --                      & \textcolor{tab20_3}{omgili} (4.78\%)           & \textcolor{tab20_16}{FacebookBot} (0.80\%)     & \textcolor{tab20_12}{Claude-Web} (9.93\%)       & \textcolor{tab20_6}{aiHitBot} (0.39\%)        \\
			7  & \textcolor{tab20_4}{magpie-crawler} (0.94\%)         & --                      & \textcolor{tab20_10}{Anthropic-AI} (3.72\%)     & \textcolor{tab20_3}{omgili} (0.80\%)          & \textcolor{tab20_13}{omgilibot} (8.48\%)        & \textcolor{tab20_8}{ChatGPT-User} (0.39\%)    \\
			8  & \textcolor{tab20_5}{Scrapy} (0.28\%)       & --                      & \textcolor{tab20_2}{PetalBot} (3.48\%)         & \textcolor{tab20_13}{omgilibot} (0.80\%)       & Cohere-AI (7.20\%)        & \textcolor{tab20_14}{PerplexityBot} (0.39\%)  \\
			9  & \textcolor{tab20_6}{aiHitBot} (0.14\%)     & --                      & \textcolor{tab20_12}{Claude-Web} (3.35\%)       & \textcolor{tab20_15}{Amazonbot} (0.80\%)       & \textcolor{tab20_15}{Amazonbot} (6.87\%)        & \textcolor{tab20_4}{magpie-crawler} (0.39\%)  \\
			10 & \textcolor{tab20_7}{ByteSpider} (0.10\%)    & --                      & \textcolor{tab20_4}{magpie-crawler} (1.19\%)   & \textcolor{tab20_4}{magpie-crawler} (0.40\%)  & \textcolor{tab20_11}{ClaudeBot} (6.51\%)        & \textcolor{tab20_5}{Scrapy} (0.19\%)         \\
			\midrule
			& \multicolumn{2}{c|}{\textbf{09/2024}} 
			& \multicolumn{2}{c|}{\textbf{01/2025}} 
			& \multicolumn{2}{c}{\textbf{05/2025}} \\
			\cmidrule(lr){2-3} \cmidrule(lr){4-5} \cmidrule(lr){6-7} 
			\# & \textbf{Reputable News}          & \textbf{Misinformation}          & \textbf{Reputable News}            & \textbf{Misinformation}           & \textbf{Reputable News}            & \textbf{Misinformation}           \\
			\midrule
			1& \textcolor{tab20_0}{GPTBot} (52.38\%)          & \textcolor{tab20_2}{PetalBot} (4.16\%)        & \textcolor{tab20_0}{GPTBot} (51.97\%)          & \textcolor{tab20_2}{PetalBot} (3.56\%)        & \textcolor{tab20_0}{GPTBot} (52.54\%)          & \textcolor{tab20_0}{GPTBot} (4.80\%)             \\
			2& \textcolor{tab20_9}{Google-Extended} (43.08\%) & \textcolor{tab20_0}{GPTBot} (2.84\%)          & \textcolor{tab20_1}{CCBot }(47.93\%)           & \textcolor{tab20_0}{GPTBot} (3.37\%)          & \textcolor{tab20_1}{CCBot }(48.31\%)           & \textcolor{tab20_15}{Amazonbot} (3.87\%)          \\
			3& \textcolor{tab20_1}{CCBot }(42.91\%)           & \textcolor{tab20_17}{SemrushBot-SWA} (1.51\%)  & \textcolor{tab20_9}{Google-Extended} (43.51\%) & \textcolor{tab20_1}{CCBot }(2.81\%)           & \textcolor{tab20_8}{ChatGPT-User} (45.25\%)    & \textcolor{tab20_2}{PetalBot} (3.32\%)           \\	
			4& \textcolor{tab20_8}{ChatGPT-User} (36.97\%)    & \textcolor{tab20_8}{ChatGPT-User} (1.13\%)    & \textcolor{tab20_8}{ChatGPT-User} (41.39\%)    & \textcolor{tab20_9}{Google-Extended} (2.62\%) & \textcolor{tab20_9}{Google-Extended} (43.69\%) & \textcolor{tab20_9}{Google-Extended} (3.14\%)    \\
			5& \textcolor{tab20_10}{Anthropic-AI} (18.94\%)    & \textcolor{tab20_1}{CCBot }(1.13\%)           & \textcolor{tab20_10}{Anthropic-AI} (39.47\%)    & \textcolor{tab20_10}{Anthropic-AI} (2.25\%)    & \textcolor{tab20_10}{Anthropic-AI} (43.60\%)    & \textcolor{tab20_1}{CCBot }(3.14\%)              \\
			6& \textcolor{tab20_13}{omgilibot} (18.16\%)       & \textcolor{tab20_9}{Google-Extended} (1.13\%) & \textcolor{tab20_14}{PerplexityBot} (38.69\%)   & \textcolor{tab20_15}{Amazonbot} (2.06\%)       & \textcolor{tab20_11}{ClaudeBot} (42.17\%)       & \textcolor{tab20_11}{ClaudeBot} (2.95\%)          \\
			7& \textcolor{tab20_14}{PerplexityBot} (18.03\%)   & \textcolor{tab20_14}{PerplexityBot} (0.95\%)   & \textcolor{tab20_11}{ClaudeBot} (38.01\%)       & \textcolor{tab20_16}{FacebookBot} (1.87\%)     & \textcolor{tab20_12}{Claude-Web} (41.78\%)      & \textcolor{tab20_7}{ByteSpider} (2.95\%)         \\
			8& \textcolor{tab20_11}{ClaudeBot} (17.18\%)       & \textcolor{tab20_15}{Amazonbot} (0.76\%)       & \textcolor{tab20_12}{Claude-Web} (37.65\%)      & \textcolor{tab20_14}{PerplexityBot} (1.50\%)   & \textcolor{tab20_13}{omgilibot} (41.48\%)       & Meta-ExternalAgent (2.58\%) \\
			9& \textcolor{tab20_12}{Claude-Web} (17.15\%)      & \textcolor{tab20_11}{ClaudeBot} (0.57\%)       & \textcolor{tab20_13}{omgilibot} (37.49\%)       & \textcolor{tab20_11}{ClaudeBot} (1.50\%)       & \textcolor{tab20_3}{omgili} (39.86\%)          & \textcolor{tab20_10}{Anthropic-AI} (2.58\%)       \\
			10& \textcolor{tab20_3}{omgili} (15.94\%)          & \textcolor{tab20_10}{Anthropic-AI} (0.57\%)    & \textcolor{tab20_16}{FacebookBot} (36.71\%)     & \textcolor{tab20_8}{ChatGPT-User} (1.50\%)    & \textcolor{tab20_14}{PerplexityBot} (39.11\%)   & \textcolor{tab20_16}{FacebookBot} (2.21\%)        \\

\bottomrule
		\end{tabular}
	}
\end{table*}


\vspace{2mm}
\begin{mdframed}[linecolor=cyan!60!black, backgroundcolor=cyan!5]
\textbf{Key takeaway:} 
Reputable news sites are 
systematically adapting to the rise of AI crawlers by expanding and updating their \robots directives. In contrast, misinformation sites remain largely passive, rarely adopting such measures and leaving their content relatively more accessible to AI agents.

\end{mdframed}

\section{Discussion}
\label{sec:Discussion}




In this section, we reflect on the broader implications of our findings, acknowledge the limitations of our methodology, and outline the ethical considerations underlying our data collection. 

\subsection{Implications}
Our results show that reputable news sites are significantly more likely to include \robots directives disallowing AI user agents, while misinformation sites rarely do. This suggests that reputable news sources are more actively asserting control over how their content is accessed by automated systems, particularly those associated with AI applications. In contrast, the absence of such directives on misinformation websites may reflect limited awareness, technical capacity, or willingness to restrict access.
This disparity in declared access policies raises concerns about which types of content are more readily available to AI crawlers. While we do not evaluate training datasets, the continued accessibility of misinformation sites may increase the likelihood that their content is collected and reused, especially as reputable news sites increasingly opt out.

In addition, the potential emergence of a self-reinforcing loop in which permissive misinformation websites, some of which may already be AI-generated themselves \cite{newsGuardAINews}, continue to allow unrestricted scraping by web crawlers. This openness facilitates the redistribution of their content across various platforms and datasets, increasing the likelihood that such misinformation will be repeatedly ingested into future training corpora. As a result, the same misleading or fabricated information may reappear in new AI-generated content, amplifying its reach and reinforcing its perceived legitimacy over time.

\subsection{Limitations}
Firstly, we follow prior work in using MBFC to classify websites~\cite{chen2020proactive,chowdhury2020joint,ghanem2021fakeflow,gruppi2021nela,zhou2020recovery,papadogiannakis2023funds}; however, we acknowledge that such labels are based on human judgment and may carry inherent subjectivity. Additionally, MBFC's coverage is focused on English-language sources and may not capture the full diversity of online media ecosystems. 
Secondly, while prior work suggests that major AI crawlers that claim to respect \robots generally comply~\cite{liu2025somesite}, we do not verify crawler behavior. 
As such, our analysis captures the intended access restrictions specified by websites, rather than the actual behavior of AI agents in practice.
In addition, our analysis focuses on \robots directives and crawler access controls; we do not have visibility into the specific websites included in proprietary LLM training datasets. Therefore, we cannot establish a direct or causal relationship between a site's exclusion policies and its presence or absence in training data. 
Finally, the \robots directives used by reputable news and misinformation websites may depend on factors such as organization size, monetization models, or the need to protect valuable content. 
We intend to systematically examine the role of these factors in shaping blocking behavior in future work.

\subsection{Ethics}
To minimize resource usage and avoid unnecessary load on websites, our crawler was configured to retrieve only the \robots file and the HTTP response status of the homepage. We did not access any additional pages or collect any content beyond these publicly accessible metadata endpoints. The crawling agent used in this study included a clearly identified \texttt{User-agent} string and provided a contact email to allow website administrators to raise concerns or request exclusion. All requests respected the REP, and no attempts were made to bypass access restrictions or interact with user-generated content. We believe this approach aligns with responsible web measurement practices, supporting transparency and reproducibility.





\section{Conclusion}
In this study, we examined how reputable and misinformation websites use \texttt{robots.txt} directives to manage access by AI crawlers. By analyzing 3,369 reputable news websites and 710 misinformation websites, we found a consistent and growing disparity in \robots exclusion behavior. As of May 2025, 60.0\% of reputable websites disallowed at least one AI agent, compared to only 9.1\% of misinformation websites. 
Even in the absence of \robots, we find active blocking is additionally used as a control mechanism to restrict AI crawler access. 
Reputable news websites also listed significantly more AI agents in their blocklists and adopted more selective and comprehensive exclusion policies over time.

Our findings highlight that reputable news websites are more actively restricting access through their \robots policies in response to AI scraping,
while misinformation websites remain relatively permissive. As a result, the content landscape accessible to AI models may shift in unintended ways. While we do not make claims about downstream model behavior, our results underscore the importance of access control mechanisms like \texttt{robots.txt} in shaping the data environment used to train large-scale AI systems.


\bibliographystyle{ACM-Reference-Format}
\bibliography{ref}

\appendix

\section{Ranking of Reputable News and Misinformation Websites}
\begin{figure}[h!]
    \centering    
    \includegraphics[width=\linewidth]{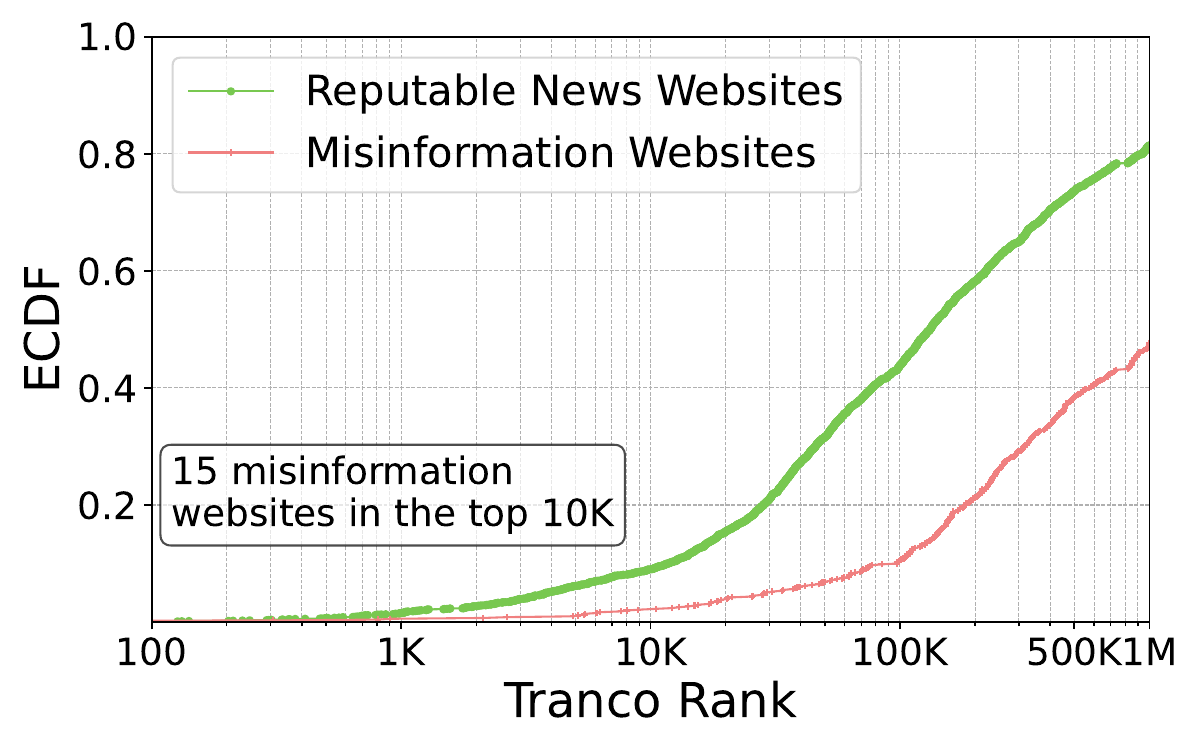}
    \caption{Distribution of rank of news websites based on Tranco. There are both reputable and misinformation websites with
very high popularity, 
including 15 misinformation websites in the Top 10K.
}
    \label{fig:rank_1}
\end{figure}
\autoref{fig:rank_1} shows the distribution of website popularity (Tranco rank) for reputable and misinformation news websites. Both groups include some highly ranked domains, with 15 misinformation sites appearing within the Top 10K.



\end{document}